\documentclass[11.5pt]{revtex4}

\usepackage{graphicx}
\usepackage{dcolumn}
\usepackage{color}
\usepackage{bm}
\usepackage{amsmath}
\usepackage{amsfonts}
\usepackage{amssymb}

\input epsf

\begin{document}

\title{A cosmographic analysis of holographic dark energy models}

\author{Supriya Pan\footnote{span@research.jdvu.ac.in}}
\affiliation{Department of Mathematics, Jadavpur University, Kolkata-700 032, India.}

\author{Subenoy Chakraborty\footnote{schakraborty@math.jdvu.ac.in}}

\affiliation{Department of Mathematics, Jadavpur University, Kolkata-700 032, India.}{}

\begin{abstract}
The present work deals with a detailed study of interacting holographic dark energy model for three common choices of the interaction term. Also, two standard choices of IR cut-off, namely, Ricci length scale and radius of the event horizon are considered here. Finally, the cosmographic parameters are presented both analytically and graphically.\\


Pacs No: 95.36.+x, 98.65.-r, 98.80.Es.

\end{abstract}
\maketitle
\section{Introduction}
Cosmology got tremendous excitement since 1998 when observations from type Ia Supernovae \cite{Riess1} demanded that our universe is accelerating. This observational prediction was also supported by cosmic microwave background radiation \cite{Komatsu1}, SDDS \cite{Tegmark1}, baryon acoustic oscillation \cite{Sanchez1}, weak lensing \cite{Jain1} etc. Standard cosmology can only explain this observational fact if the cosmic fluid in recent past is dominated by exotic matter having large negative pressure. Further, the observations predict that nearly 73\% of our universe is filled up with that type of components dubbed dark energy (DE). The simplest choice for the dark energy candidate is the cosmological constant $\Lambda$ and the favored cosmological model which fits most of the observational data is the $\Lambda$--cold--dark--matter ($\Lambda$CDM) model which represents a vacuum energy density having equation of state parameter (EoS) $\omega= -1$. Although the model predicts cosmic acceleration as well as a reasonable agreement with observational data, there are some embarrassing issues related to this model, namely, cosmological constant problem \cite{Weinberg1, Carroll1} (the huge discrepancy between the observed value of the cosmological constant and the one predicted in quantum field theory), coincidence problem \cite{Copeland1} (although generically small, but the cosmological constant happens to be exactly of the value required to become dominant at the present epoch) and, recently, it was shown that the $Λ$CDM model may also suffer from the age problem \cite{Yang1}. Due to those problems in the above model, scalar field models, namely, quintessence \cite{Caldwell1}, phantom \cite{Caldwell2}, K--essence \cite{Armendariz-Picon1}, Tachyon \cite{Padmanabhan1}, Quintom \cite{Elizalde1} attracted special attention as dynamical DE models.\\

At present DE and cold dark matter (CDM, thus dark matter will be written as DM) are dominant sources of our universe. So, it is very natural to consider the interaction between the two dark sectors DE and CDM \cite{Ma1}. The interaction between the two dark sectors will give a richer dynamics than the non-interacting case. Also, it has been shown that an appropriate choice of the interaction between DE and DM can alleviate the cosmic coincidence problem \cite{Pavon1}.\\

In 2004, a new model of dark energy was introduced by Miao Li \cite{Li1}, based on the holoraphic principle. According to the holographic principle, the number of degrees of freedom in a bounded system should be finite and related to the area of its boundary and based on this principle, a field theoretic relation between short distance (ultraviolet) cut off and a long distance (infra red) cut off were established \cite{Cohen1}. This relation ensures that the energy in a box of size L does not exceed the energy of a black hole of the same size. The general expression for the energy density of a holographic dark energy is (choosing 8$\pi$G= 1) $$\rho_d= \frac{3c^2}{L^2},$$
where, the dimensionless parameter $c^2$ takes care of the uncertainties of the theory, L is the IR cut-off and the factor of three has been introduced for mathematical convenience. In the present work, we choose L = $L_R$, the Ricci length or L = $R_E$, the radius of the future event horizon as the IR cut off in two different sections. The argument behind the choice of the Ricci length \cite{Gao1} as IR cut off is that it corresponds to the size of the maximal perturbation, leading to the formation of a black hole \cite{Burstein1}. On the other hand, radius of the future event horizon is commonly used as the IR cut-off of HDE models which gives the correct equation of state and the desired accelerating universe. However, recently it has been shown \cite{Duran1} that future event horizon suffers from a severe circularity problem.\\

On the other hand, the interaction between DE and DM could led a major issue to be confronted in studying the physics of DE. However, due to the nature of these two components remaining unknown, it will not be possible to derive the precise form of the interaction from the outset or determine it from phenomenological requirements. Further, in the framework of field theory \cite{Pavon1}, it is natural to consider the inevitable interaction between the dark components. An appropriate interaction between the DE and DM can provide a mechanism to alleviate the coincidence problem. Moreover, in view of the continuity equations (see Eq. (4) in Section II), the interaction between DE and DM must be a function of the energy densities multiplied by a quantity with units of the inverse of time which has the natural choice as Hubble parameter. Thus interaction between DE (energy density $\equiv$ $\rho_d$) and DM (energy density $\equiv$ $\rho_m$) could be expressed phenomenologically in forms such as (i) $Q= Q(H \rho_m)$, (ii) $Q= Q(H \rho_d)$, (iii) $Q= Q[H(\rho_d+ \rho_m)]$, or more generally, (iv) $Q= Q(H \rho_d, H \rho_m)$. In the present work, for simplicity the interaction terms are chosen as follows:\\

(I) $3b^2 H(\rho_m+ \rho_d)$---a linear combination of the energy densities of DE and DM components.\\

(II) a natural and physically viable interaction term of the form $\frac{\gamma}{H}$ $\rho_m \rho_d$, where, $\gamma$ is a constant, and\\

(III) another frequently used interaction in literature of the form $3 \lambda H \rho_d$, $\lambda$ is a constant.\\

In 2003, Sahni et al. \cite{Sahni1} proposed state finder parameters $r$, $s$ which are defined as $$r=\frac{1}{aH^3}\dddot{a},~~and~~s= \frac{r-1}{3(q-\frac{1}{2})},$$

where, $`a$' is the scale factor of the universe for the FRW model and $H$ and $q$ (= $-\frac{a \ddot{a}}{\dot{a}^2}$) are the Hubble parameter and the deceleration parameter respectively. In fact the parameter `$r$' forms the next step in the hierarchy of the geometrical cosmological parameters after $H$ and $q$. These dimensionless parameters characterize the properties of dark energy in a model independent manner. According to Sahni et al. \cite{Sahni1}, trajectories in the $r$, $s$ plane corresponding to different cosmological models demonstrate qualitative different behavior, and therefore, the state finder diagnostic together with observations may discriminate between different DE models. Inspired by the above work, some more general geometrical cosmological parameters are introduced. In fact, they are obtained from the Taylor series expansion of the scale factor. These geometrical cosmological parameters are the jerk parameter $j$ (same as $r$), snap parameter $s$ (different from the above defined $s$ parameter by Sahni et al. \cite{Sahni1}), lerk parameter $l$ and $m$ parameter. The study of the above four parameters for a particular dark energy model together with the deceleration parameter is known as the Cosmography of the model. In this connection, one may see the evolution equations for the modified and the interacting modified holographic Ricci dark energy and their state finder diagnoses \cite{Mathew1}. In the present work, we study the cosmographic analysis for the interacting DE model for the above three choices of the interaction term.\\

It is interesting to mention that due to several theoretical problems, in most of the holographic dark energy models, it is difficult to realize the equation of state parameter across `$-1$', but in this study, it is shown that the present interacting DE model can describe the evolution of the universe across the cosmological constant boundary $\omega= -1$. The paper is organized as follows: Basic equations and HDE at Ricci scale is presented in section II. Section III contains a discussion on HDE at future event horizon. In Section IV, we have introduced the cosmographic parameters (CP) and shown the variation of the CP for the HDE at Ricci scale for three above different interactions. Finally, the summary of the whole work is presented in Section V.

\section{Interacting holographic dark energy at Ricci scale}

Let the cosmic substratum is made of pressureless dark matter ({\it i.e.,} CDM) and a holographic dark energy (HDE) at Ricci scale and the two dark components are interacting in nature. In the background of the flat FRW universe, the field equations take the forms (choosing 8$\pi G = 1$)

\begin{equation}
3H^2 = \rho_ m +\rho_ d,~~~~~~~~~~~~ and~~~~~~~~~~2\dot{H}+ 3H^2= -p_d,
\end{equation}

where, $\rho_ m$ is the energy density of DM in the form of dust while ($\rho_d$, $p_d$) are the energy density and the thermodymanic pressure of the HDE in the form of a perfect fluid having equation of state, $p_d= \omega_d \rho_d$ ($\omega_d$, a variable). In the present model, as $$L = L_R = (\dot{H}+2 H^2)^{-\frac{1}{2}},$$ so,

\begin{equation}
\rho_d = 3c^2 (\dot{H}+2 H^2),
\end{equation}

and consequently the expression for the equation of state parameter becomes

\begin{equation}
\omega_d = -\frac{2}{3c^2}+\frac{1}{3 \Omega_d},
\end{equation}

where, $\Omega_d = \frac{\rho_d}{3 H^2}$ is the density parameter for the HDE. The conservation equations of the energy densities of the dark components are

\begin{equation}
\dot{\rho}_m+ 3H p_m = Q,~~~~~~~~~~~~and~~~~~~~~~~~~~~~\dot{\rho}_d+ 3H(\rho_ d+ p_d)= -Q.
\end{equation}

Here, $Q> 0$ is the indication of the flow of energy from DE to CDM and $Q< 0$ just represents the flow of energy in the opposite direction.\\

At first if we neglect the interaction ({\it i.e.,} $Q= 0$), then the evolution of the density parameter takes the form

\begin{equation}
\dot{\Omega}_d = H (1-\Omega_d)(1-\frac{2 \Omega_d}{c^2}),
\end{equation}

or, using $x$ = $ln$a, {\it i.e.,} $\frac{d}{dx} = \frac{1}{H} \frac{d}{dt}$, the above equation becomes

\begin{equation}
\frac{d\Omega_d}{dx} = (1-\Omega_d)(1-\frac{2 \Omega_d}{c^2}),
\end{equation}

which on integration gives

\begin{equation}
\Omega_d = \frac{c^2+2}{4}+ A_0 \left[\frac{1+e^{(\frac{4A_0}{c^2})(x+k_0)}}{1-e^{(\frac{4A_0}{c^2})(x+k_0)}}\right],
\end{equation}

where, $k_0$ is the integration constant and $$A_0= \left(\frac{c^2+2}{4}\right)^2- \frac{c^2}{2}.$$

Now in the following, we shall introduce the interaction:\\

\textbf{I.} $Q= 3H b^2 (\rho_ m+ \rho_d)$\\

Using both the conservation equations, the evolution equation for $\Omega_d$ is given by

\begin{equation}
\frac{d\Omega_d}{dx} = -\left [(1-\Omega_d)(1-\frac{2 \Omega_d}{c^2})+3b^2\right].
\end{equation}

The solution gives

\begin{equation}
\Omega_d =\frac{c^2 +2}{4}+ M \left [\frac{1-e^{2M(-x+b_0)}}{1+e^{2M(-x+b_0)}}\right],
\end{equation}

where, $b_0$ is the constant of integration and $$M^2= \left(\frac{c^2+2}{4}\right)^2-(3b^2 +1)\frac{c^2}{2}.$$

The deceleration parameter ($q$) is given by $$q= 1-\frac{\Omega_d}{c^2}.$$

\begin{figure}
\begin{minipage}{0.4\textwidth}
\includegraphics[width= 1.0\linewidth]{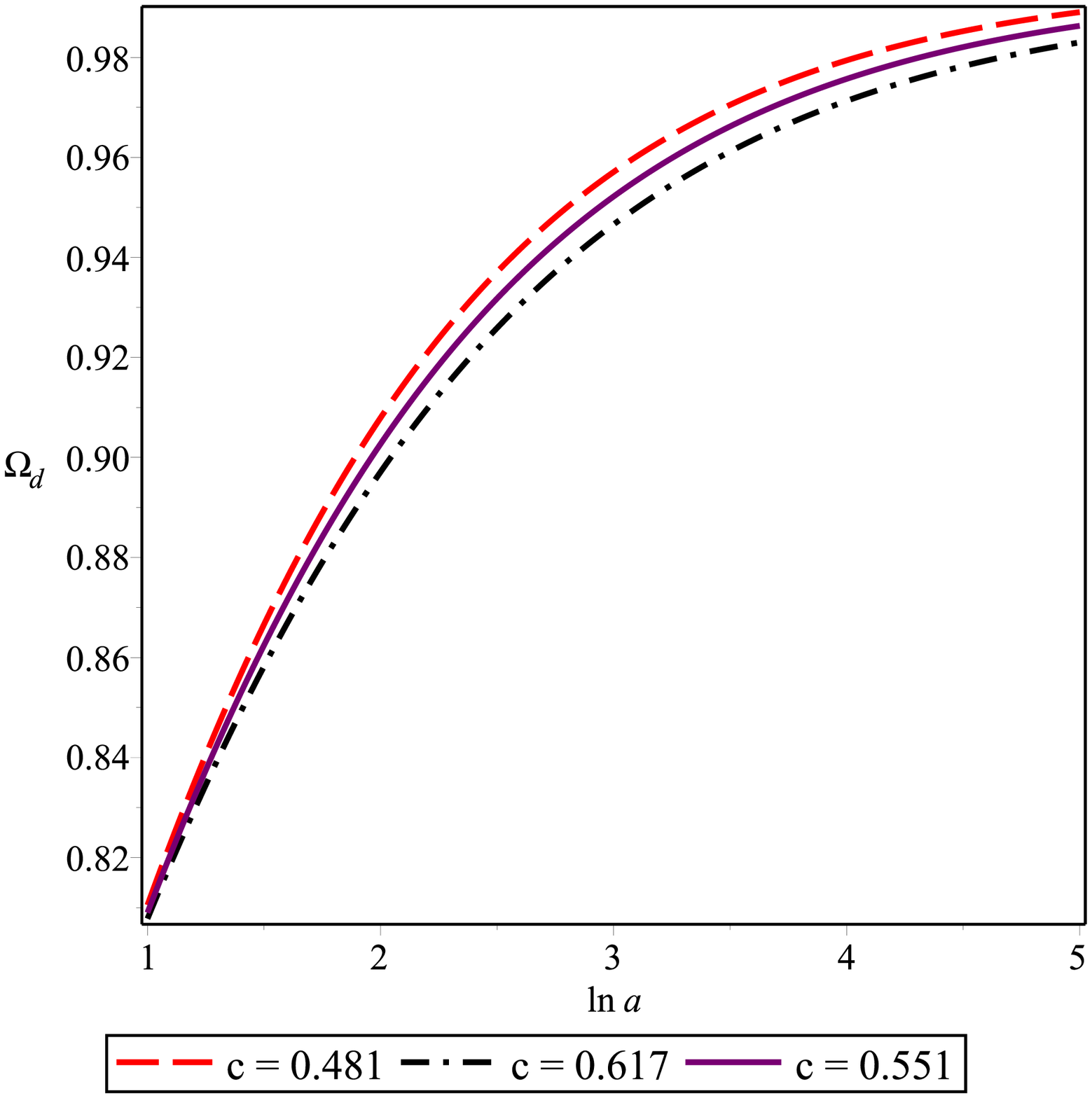}
Figure 1(a): Graphical representations of the dark energy density parameter are shown for the 1st interaction. We have used the recent Planck data for c. The coupling parameter $b^2$($= 0.01$) is taken very small.
\end{minipage}
\begin{minipage}{0.4\textwidth}
\includegraphics[width= 1.0\linewidth]{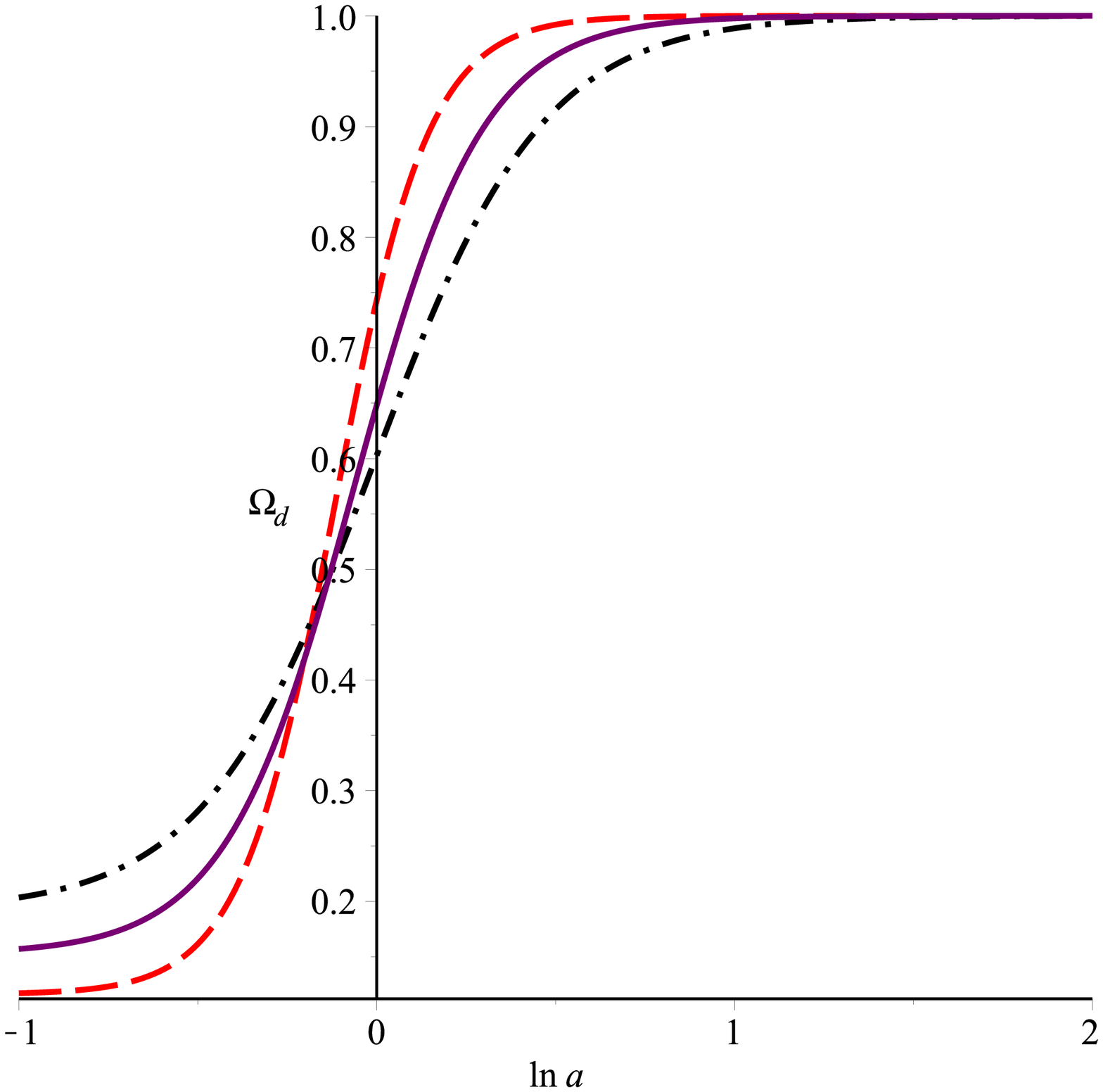}
Figure 1(b): The above graphs represent the dark energy density parameter for the second interaction with small coupling parameter $\gamma$ ($= 0.01$).
\end{minipage}
\begin{minipage}{0.4\textwidth}
\includegraphics[width= 1.0\linewidth]{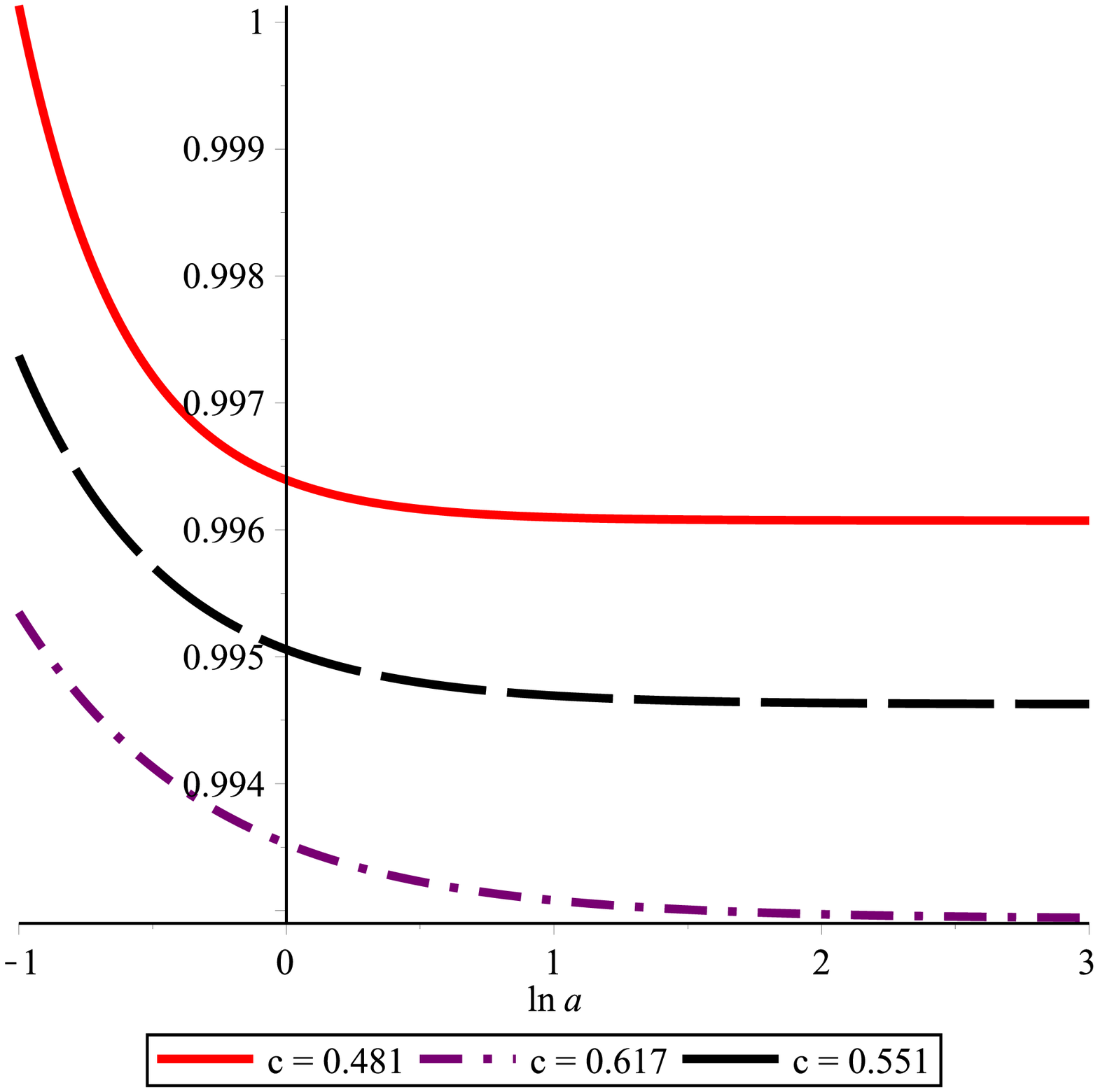}
Figure 1(c):  For the third interaction, these graphs represent the variation of the dark energy density parameter. We have taken the coupling parameter $\lambda$ ($= 0.01$) very small.
\end{minipage}
\end{figure}

\textbf{II.} $Q= \frac{\gamma}{H} \rho_m \rho_d$.\\

For this choice of the interaction term, the differential equation for the evolution of the density parameter ($\Omega_d$) has the form

\begin{equation}
\frac{d\Omega_d}{dx} = -(1-\Omega_d)\left [(1-\frac{2 \Omega_d}{c^2})+3 \gamma \Omega_d \right]
\end{equation}

and the solution for $\Omega_d$ turns out to be

\begin{equation}
\Omega_d = \frac{1+e^{(A-1)(x-a_0)}}{A+e^{(A-1)(x-a_0)}},
\end{equation}

where, $a_0$ is the constant of integration and $$A= \frac{2}{c^2}-3\gamma.$$

The deceleration parameter has the expression $q= 1- \frac{\Omega_d}{c^2}$.\\

\textbf{III.} $Q= 3 \lambda H \rho_d$.\\

For this choice of the interaction, the differential equation for the evolution of the density parameter ($\Omega_d$) has the form

\begin{equation}
\frac{d\Omega_d}{dx}= -\Omega_d \left[\left(\lambda-\frac{1}{3}-\frac{2}{3c^2}\right)+\frac{1}{3\Omega_d}+\frac{2 \Omega_d}{3c^2}\right].
\end{equation}

The solution gives

\begin{equation}
\Omega_d= -\frac{3c^2}{4}\left(\lambda-\frac{1}{3}-\frac{2}{3c^2}\right)+\left[\frac{1+e^{2E(-\frac{2x}{3c^2}+k_0)}}{1-e^{2E(-\frac{2x}{3c^2}+k_0)}}\right],
\end{equation}

where, $k_0$ is the constant of integration and

$$E= \sqrt{\frac{9c^4}{16}\left(\lambda-\frac{1}{3}-\frac{2}{3c^2}\right)^2-\frac{c^2}{2}}.$$

\begin{figure}
\begin{minipage}{0.4\textwidth}
\includegraphics[width= 1.0\linewidth]{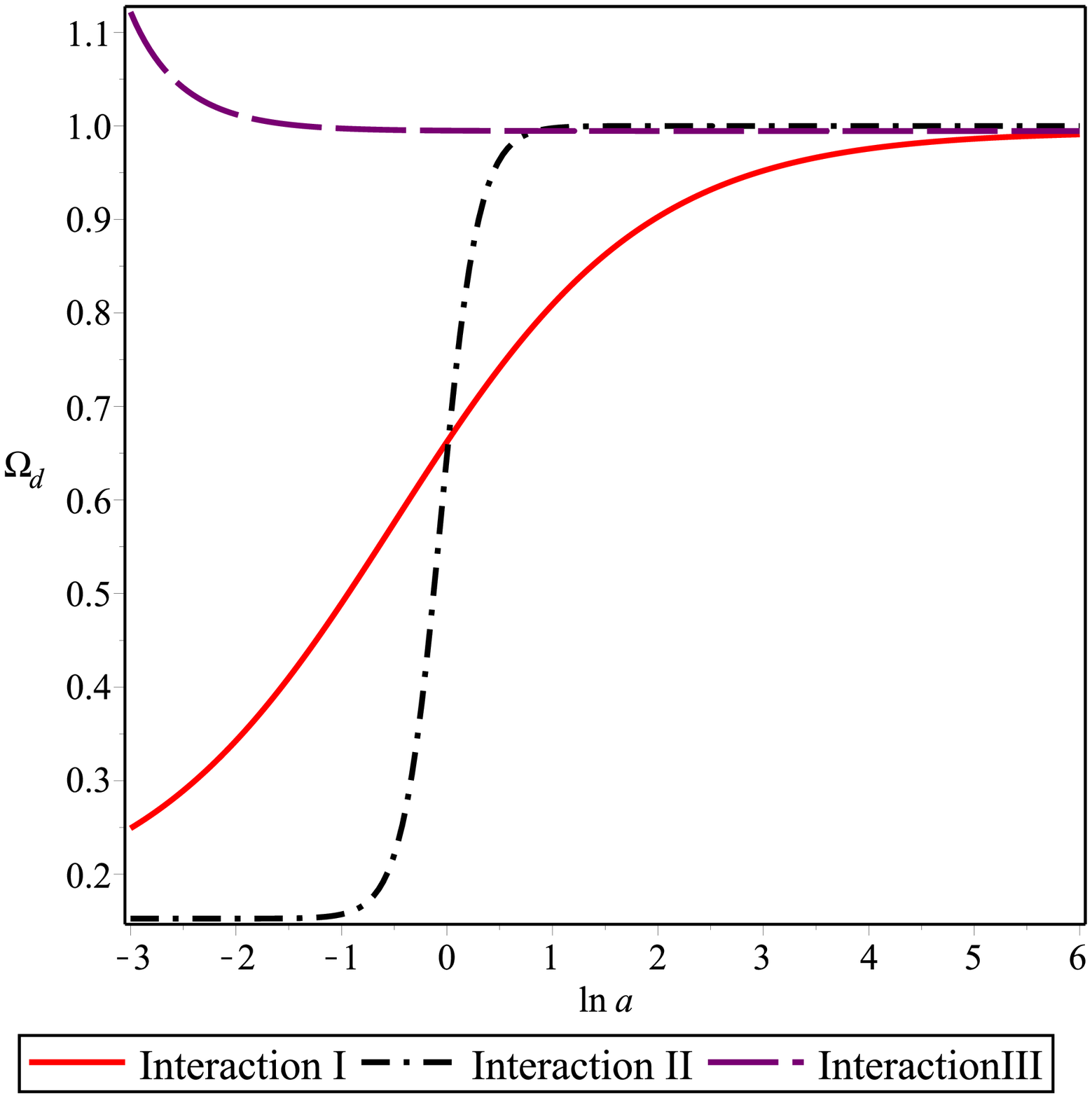}\\
Figure 1(d): For three interactions we have shown the variation of the density parameter with small coupling parameters, $b^2= \gamma= \lambda= 0.01$ and $c= 0.551$.
\end{minipage}
\end{figure}

\begin{figure}
\begin{minipage}{0.4\textwidth}
\includegraphics[width= 1.0\linewidth]{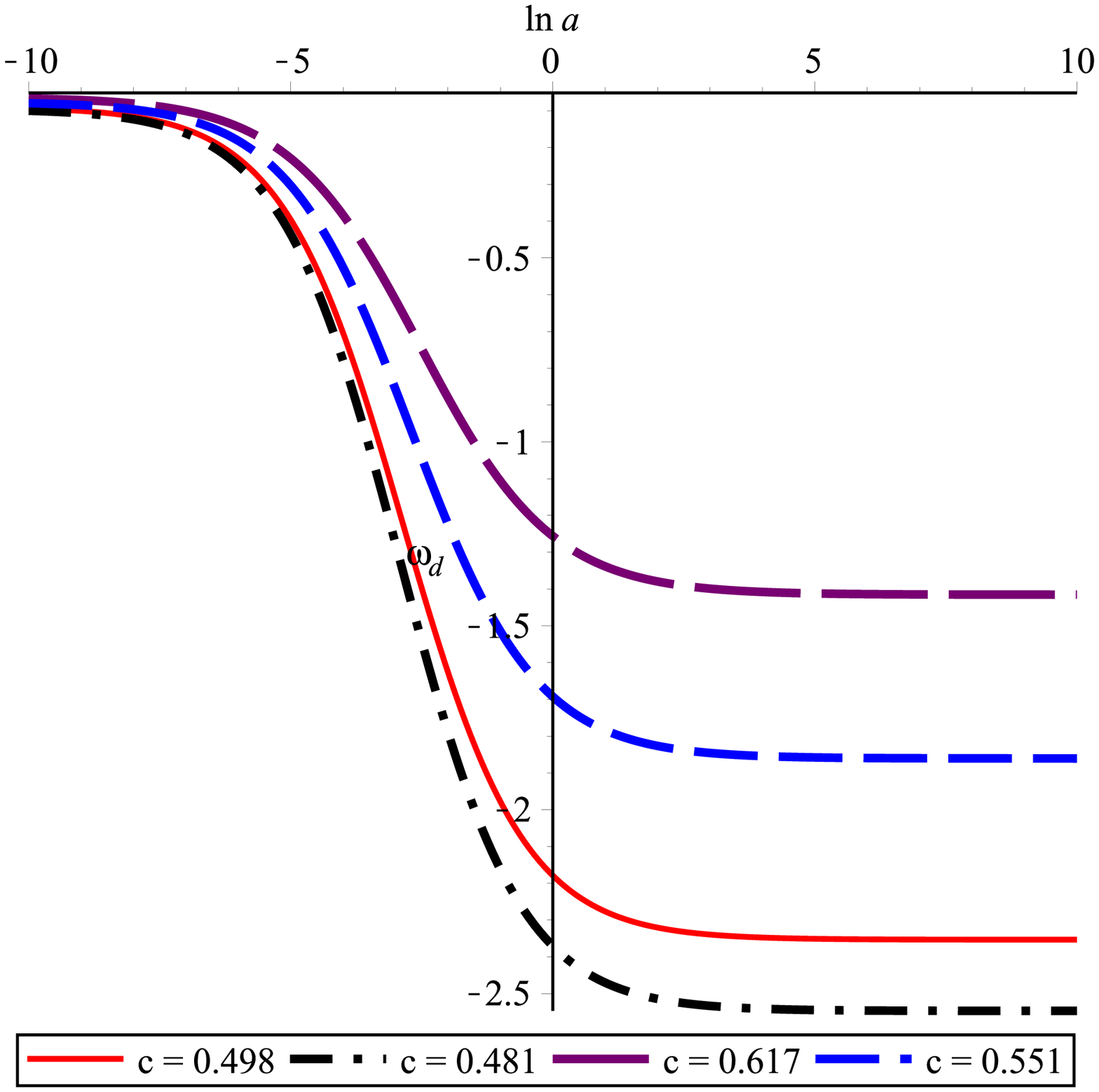}
Figure 2(a): For the first interaction, the EoS parameter for HDE are shown for three different Planck data. We have taken  very small coupling parameter ($b^2= 0.01$).
\end{minipage}
\begin{minipage}{0.4\textwidth}
\includegraphics[width= 1.0\linewidth]{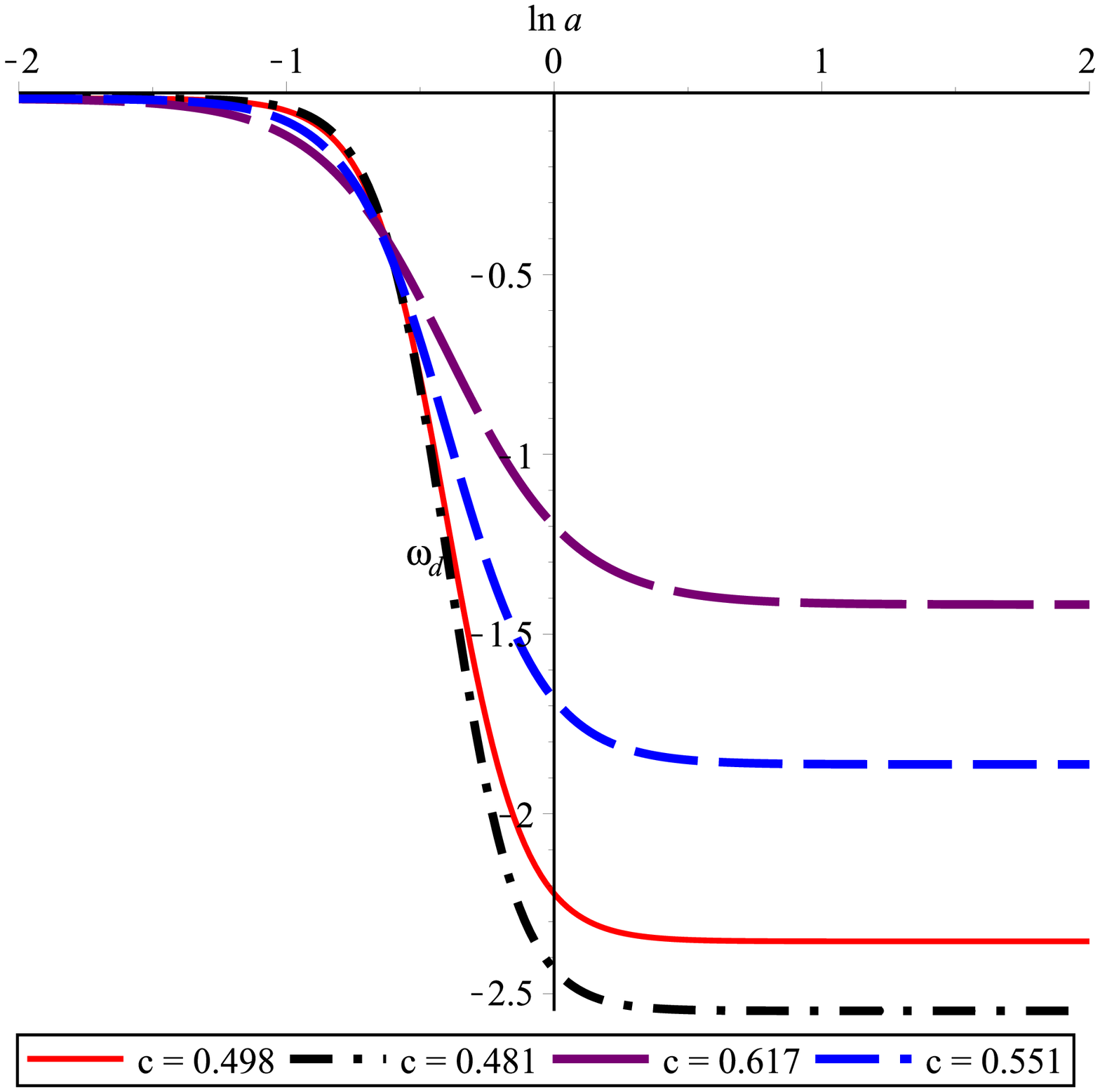}\\
Figure 2(b):  For the second interaction, the EoS parameter for HDE are shown for three different Planck data. We have taken very small coupling parameter ($\gamma= 0.01$).
\end{minipage}
\begin{minipage}{0.4\textwidth}
\includegraphics[width= 1.0\linewidth]{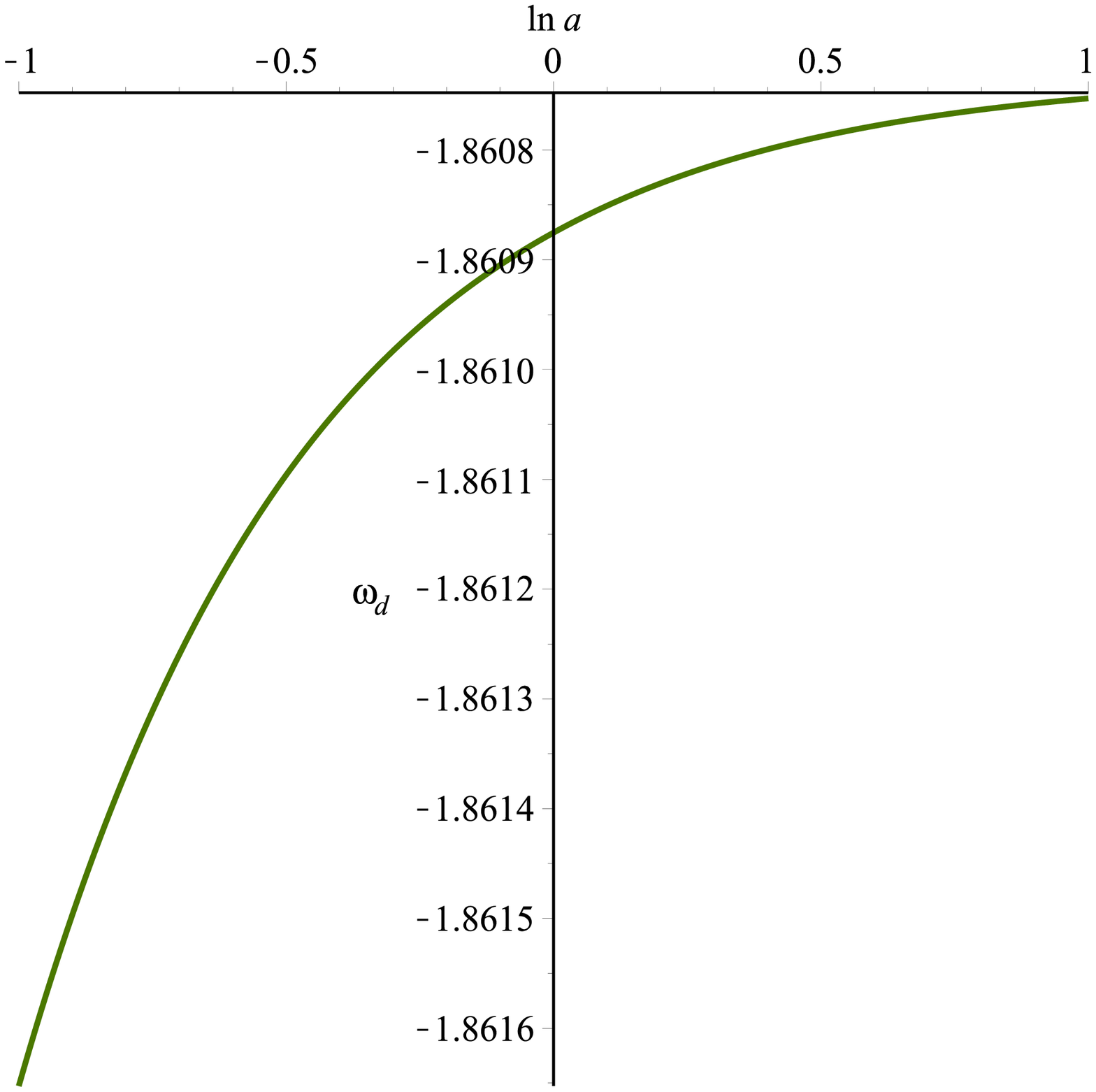}\\
Figure 2(c):  For the third interaction with small coupling parameter ($\lambda= 0.01$), the EoS parameter for HDE is plotted for the data set Planck+ WP+ Union 2.1+ BAO+ HST+ lensing.
\end{minipage}
\end{figure}

\begin{figure}
\begin{minipage}{0.4\textwidth}
\includegraphics[width= 1.0\linewidth]{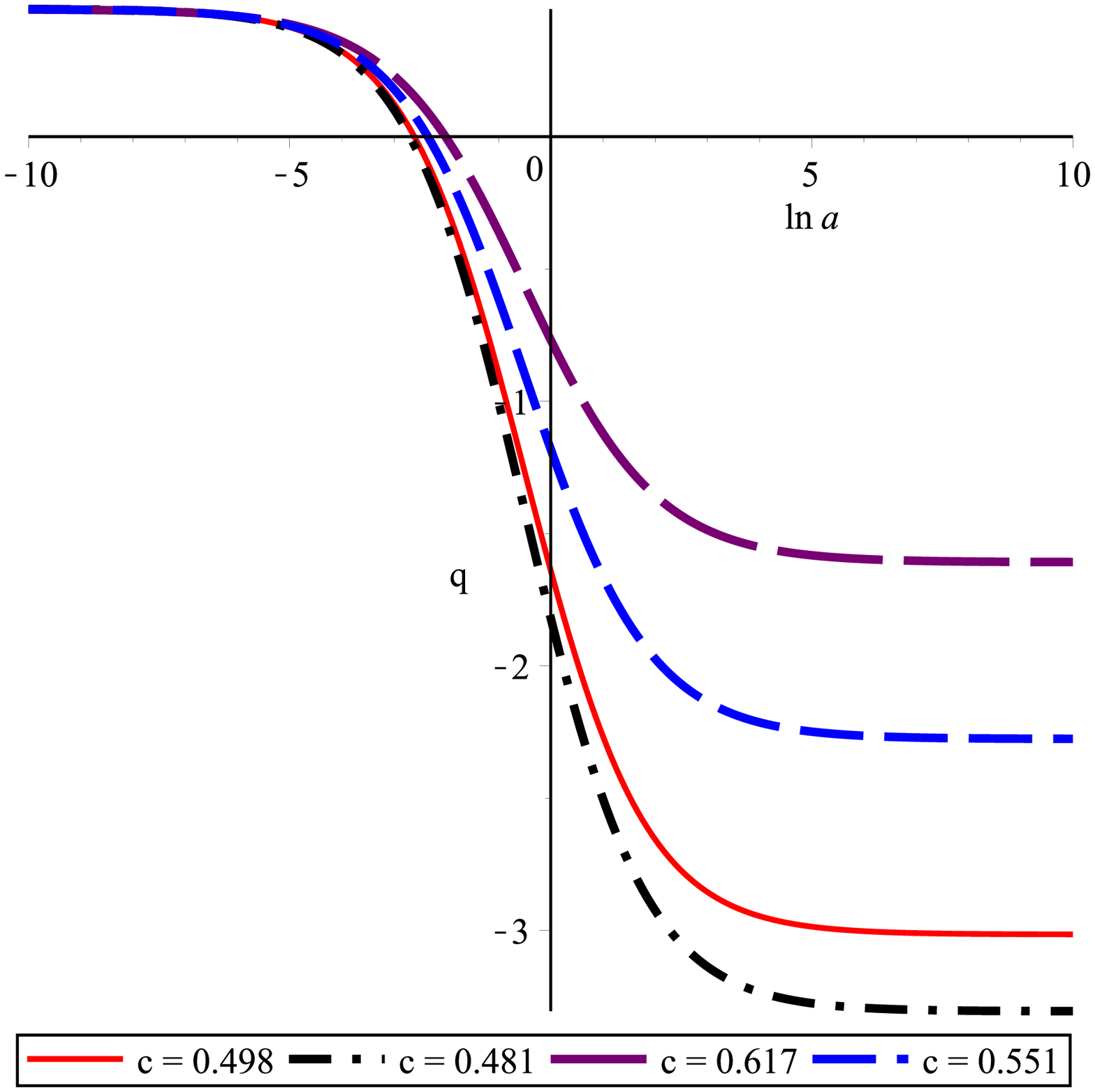}
Figure 3(a): The deceleration parameter is plotted for the first interaction with three different Planck data set. Here, $b^2= 0.01$.
\end{minipage}
\begin{minipage}{0.4\textwidth}
\includegraphics[width= 1.0\linewidth]{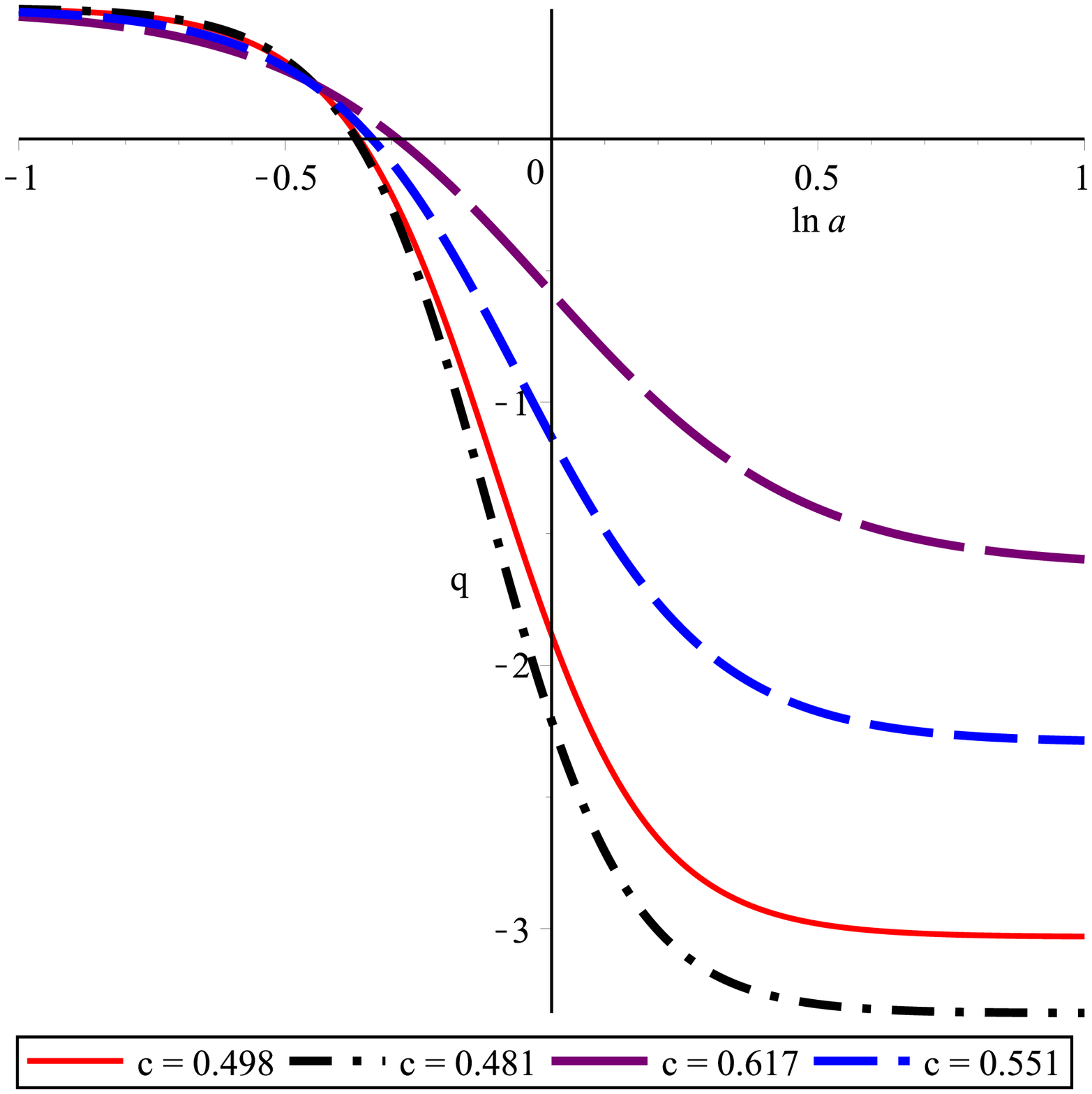}\\
Figure 3(b):  The deceleration parameter is plotted for the second interaction with three different Planck data set, where, $\gamma= 0.01$.
\end{minipage}
\begin{minipage}{0.4\textwidth}
\includegraphics[width= 1.0\linewidth]{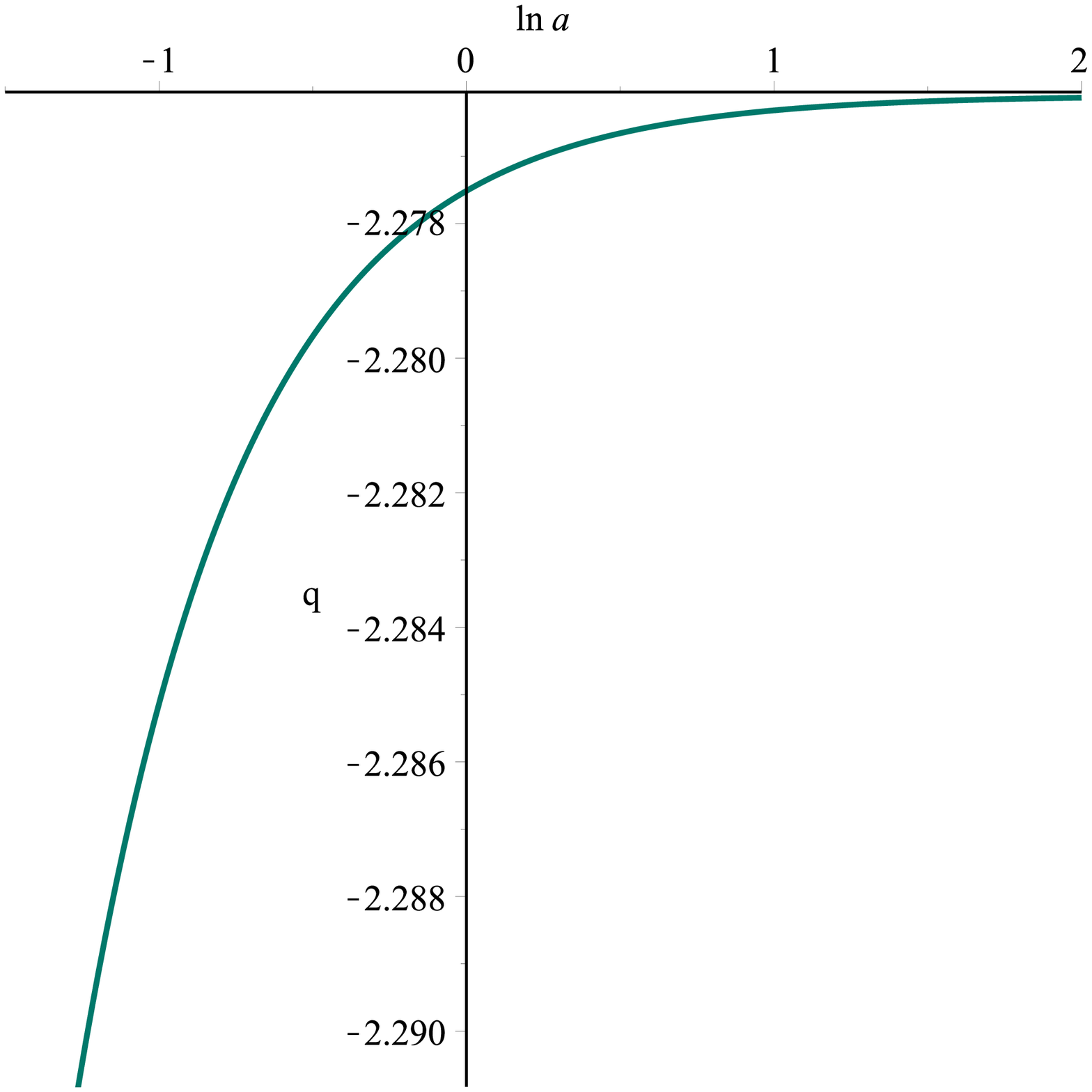}\\
Figure 3(c): For the third interaction, the deceleration parameter is shown graphically for the data set Planck+ WP+ Union 2.1+ BAO+ HST+ lensing with very small coupling parameter $\lambda= 0.01$.
\end{minipage}
\end{figure}

In Figures 1(a)--1(c); 2(a)--2(c) and 3(a)--3(c), we have shown the graphical representations for the dark energy density parameter ($\Omega_d$) for HDE, EoS parameter ($\omega_d$) and the deceleration parameter ($q$), respectively, for three different type of interactions. We have chosen the parameter $c$ from the 4 sets of recent observed Planck data \cite{Li2}. Figure 1(d) shows a comparative study of the dark energy density parameter for three different interaction terms.\\

In connection with the numerical plots, it should be mentioned that the graphs for $c= 0.498$ and $c= 0.481$ nearly coincide in Figures 1(a) and 1(b), so for clarity in the figures, we have taken only 3 data sets for $c$ \cite{Li2} and also it should be noted that while drawing the Figures 2(c) and 3(c) ({\it i.e.,} for interaction III) we have used the value of $c$ from ``Planck+ WP+ Union 2.1+ BAO+ HST+ lensing" \cite{Li2}.

\section{Interacting HDE at future event horizon}

The energy density of the dark energy has the form

\begin{equation}
\rho_d = \frac{3c^2}{R_E ^2},
\end{equation}

where $R_E$ the radius of the event horizon has the form

\begin{equation}
R_E= a \int _t ^ \infty \frac{dt}{a}.
\end{equation}

Now, for this choice of $\rho_d$, the equation of state parameter is characterized by

\begin{equation}
\omega_d= -\frac{1}{3}-\frac{2\sqrt{\Omega_d}}{3c}-\frac{b^2}{\Omega_d},~~~~~~~~~~~~~~~~~~~~when~~Q= 3Hb^2 (\rho_m+\rho_d);
\end{equation}

\begin{equation}
\omega_d= -\frac{1}{3}-\frac{2\sqrt{\Omega_d}}{c}-\gamma (1-\Omega_d),~~~~when~~Q= \frac{\gamma}{H}\rho_m\rho_d;
\end{equation}

and

\begin{equation}
\omega_d= -\frac{1}{3}-\frac{2}{3c}\sqrt{\Omega_d}-\lambda,~~~~~~~~~~~for~~Q= 3H \lambda \rho_d.
\end{equation}

The evolution of the density parameter is characterized by

\begin{equation}
\frac{d\Omega_d}{dx}= \Omega_d (1-\Omega_d)\left[1+\frac{2\sqrt{\Omega_d}}{c}-\frac{3b^2}{1-\Omega_d}\right],~~~~~for~~Q= 3Hb^2 (\rho_m+\rho_d);
\end{equation}

\begin{equation}
\frac{d\Omega_d}{dx}= \Omega_d (1-\Omega_d)\left[1+\frac{\sqrt{2\Omega_d}}{c}-3 \gamma \Omega_d \right],~~~~for~~Q= \frac{\gamma}{H} \rho_m \rho_d;
\end{equation}

and

\begin{equation}
\frac{d\Omega_d}{dx}= \Omega_d \left[(1-\Omega_d)(1+\frac{2\sqrt{\Omega_d}}{c})-3 \lambda \Omega_d \right],~~~~for~~Q= 3H \lambda \rho_d.
\end{equation}

Note that due to complicated form of the evolution equations for $\Omega_d$ for all the interactions we are unable to find an analytic expression for $\Omega_d$. Now, the expressions for the deceleration parameter for three interactions are given by

\begin{equation}
q= 1-\frac{3b^2}{2}-\frac{\Omega_d}{2}-\frac{\Omega_d ^{\frac{3}{2}}}{c},~~~~~~~~~~~~~~~~~~~~~~~~~~~~~for~~Q= 3Hb^2 (\rho_m+\rho_d);
\end{equation}

\begin{equation}
q= 1-\frac{3}{2}\Omega_d \left[\frac{1}{3}+\frac{2\sqrt{\Omega_d}}{3c}+\gamma (1-\Omega_d)\right],~~~for~~Q= \frac{\gamma}{H} \rho_m \rho_d;
\end{equation}

and

\begin{equation}
q= \frac{1}{2}-\frac{\Omega_d}{2}-\frac{\sqrt{\Omega_d}}{c}-\frac{3\lambda}{2}\Omega_d,~~~for~~Q= 3H \lambda \rho_d.
\end{equation}

In Figures 4(a)--4(c), we have plotted the variation of $\omega_d$ against $\Omega_d$ for three type of interactions. In each figure, we have taken 4 sets of Planck data for $c$ \cite{Li2}.\\

Similarly, in Figures 5(a)--5(c), the variation of $q$ over $\Omega_d$ are presented for the possible interaction with small coupling parameters. Here, we have taken 3 data sets for $c$ because the graphs for $c= 0.498$ and $c= 0.481$ almost coincide with each other. So, for the clarity in the figure, only 3 data sets are taken.\\

It is worth mentioning that due to several theoretical reasons, most of the HDE models are difficult to realize the equation of state parameter across the cosmological constant boundary $\omega= -1$. However, it is shown in the present study that the dark energy models realize such an interesting phenomenon (\cite{Elizalde1} and also for a comprehensive review see Ref. \cite{yfcai1}). Further, in this context, it should be mentioned that based on the cosmological perturbation theory, a no-go theorem (which forbids the equation of state parameter of a single perfect fluid or a single scalar field to cross the cosmological constant boundary $\omega= -1$) was discussed in Ref. \cite{yfcai2} and the corresponding proof was given in Ref. \cite{Xia1}.\\

Figures 2, 4(a) and 4(c) show such transition across `$-1$' is possible. In particular, this interesting feature in Figures 2 is very much similar to a dark energy model of a spinor field discussed by the authors in Ref. \cite{yfcai3}.

\begin{figure}
\begin{minipage}{0.4\textwidth}
\includegraphics[width= 1.0\linewidth]{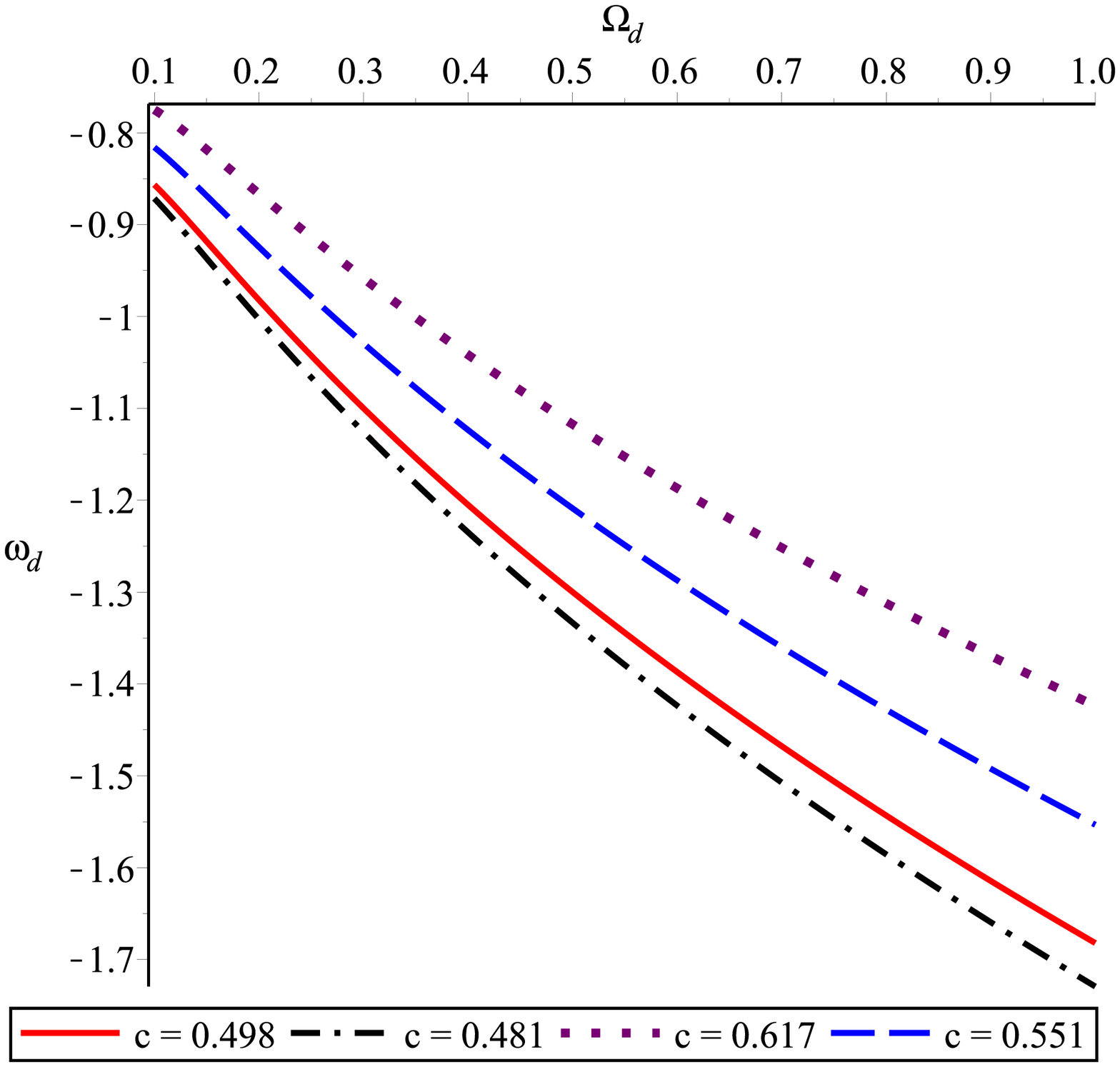}\\
Figure 4 (a):  For the holographic dark energy at future event horizon, the above three graphs represent the EoS parameter for the first interaction term with the dark energy density parameter; where $b^2= 0.01$.
\end{minipage}
\begin{minipage}{0.4\textwidth}
\includegraphics[width= 1.0\linewidth]{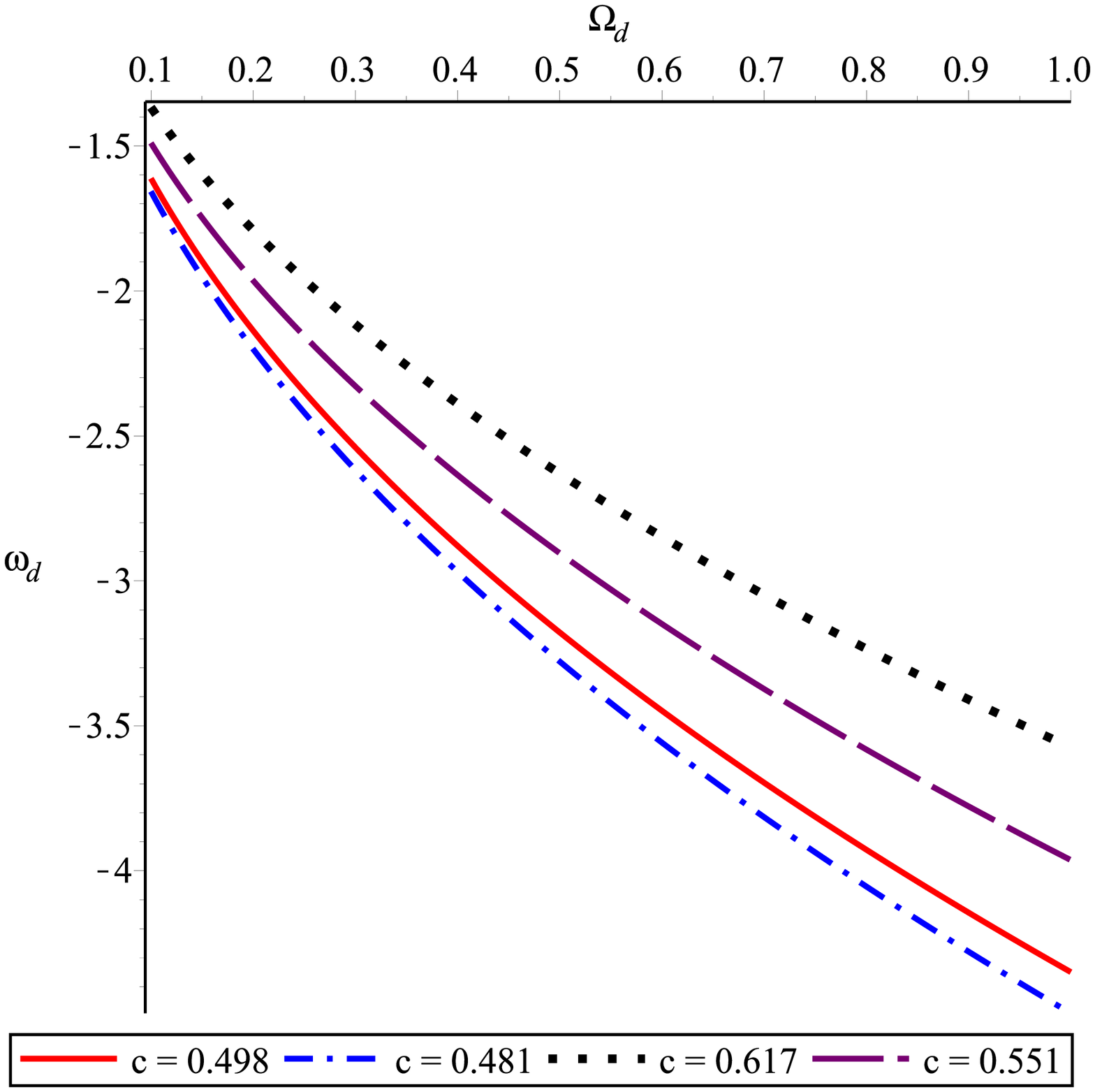}\\
Figure 4 (b):  For the second interaction in case of holographic dark energy at future event horizon, the EoS parameter is shown for three different Planck data set, where, $\gamma= 0.01$.
\end{minipage}
\begin{minipage}{0.4\textwidth}
\includegraphics[width= 1.0\linewidth]{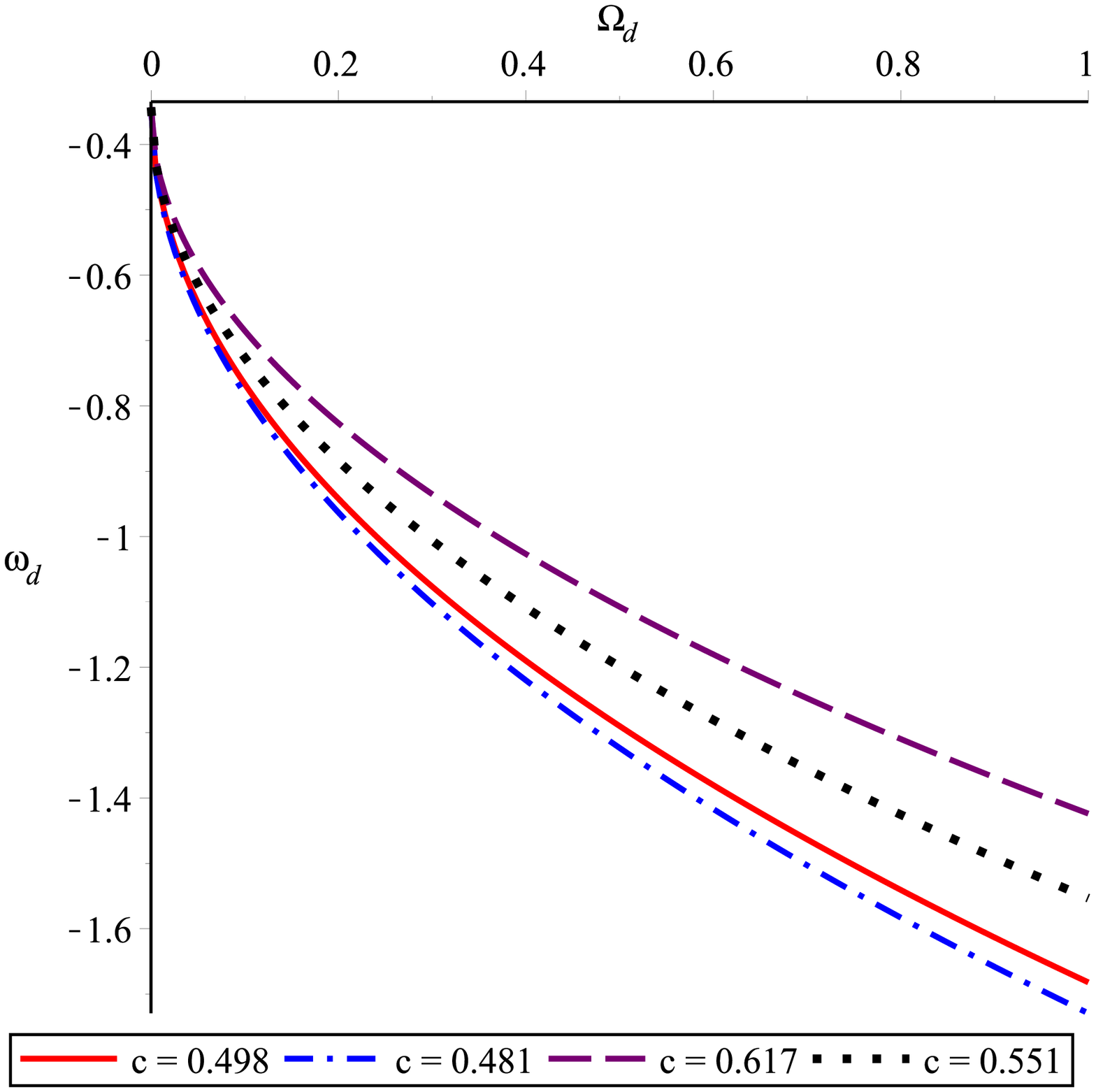}\\
Figure 4 (c):  For the third interaction in case of holographic dark energy at future event horizon, the EoS parameter is shown for three different Planck data set. Here, $\lambda= 0.01$.
\end{minipage}
\end{figure}

\begin{figure}
\begin{minipage}{0.4\textwidth}
\includegraphics[width= 1.0\linewidth]{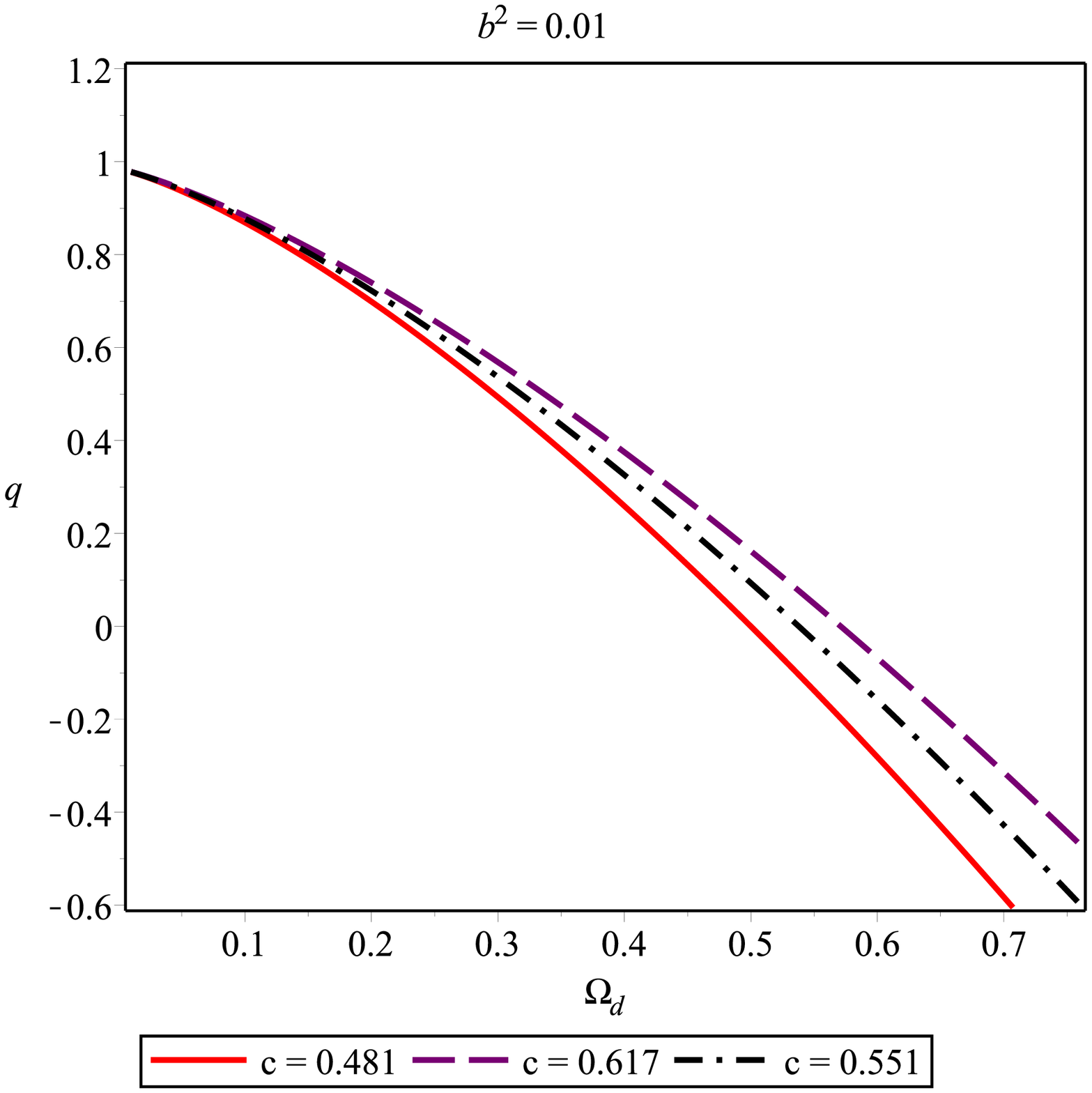}\\
Figure 5(a): Deceleration parameter versus dark energy density parameter for HDE at future event horizon is shown in three different graphs using Planck data.
\end{minipage}
\begin{minipage}{0.4\textwidth}
\includegraphics[width= 1.0\linewidth]{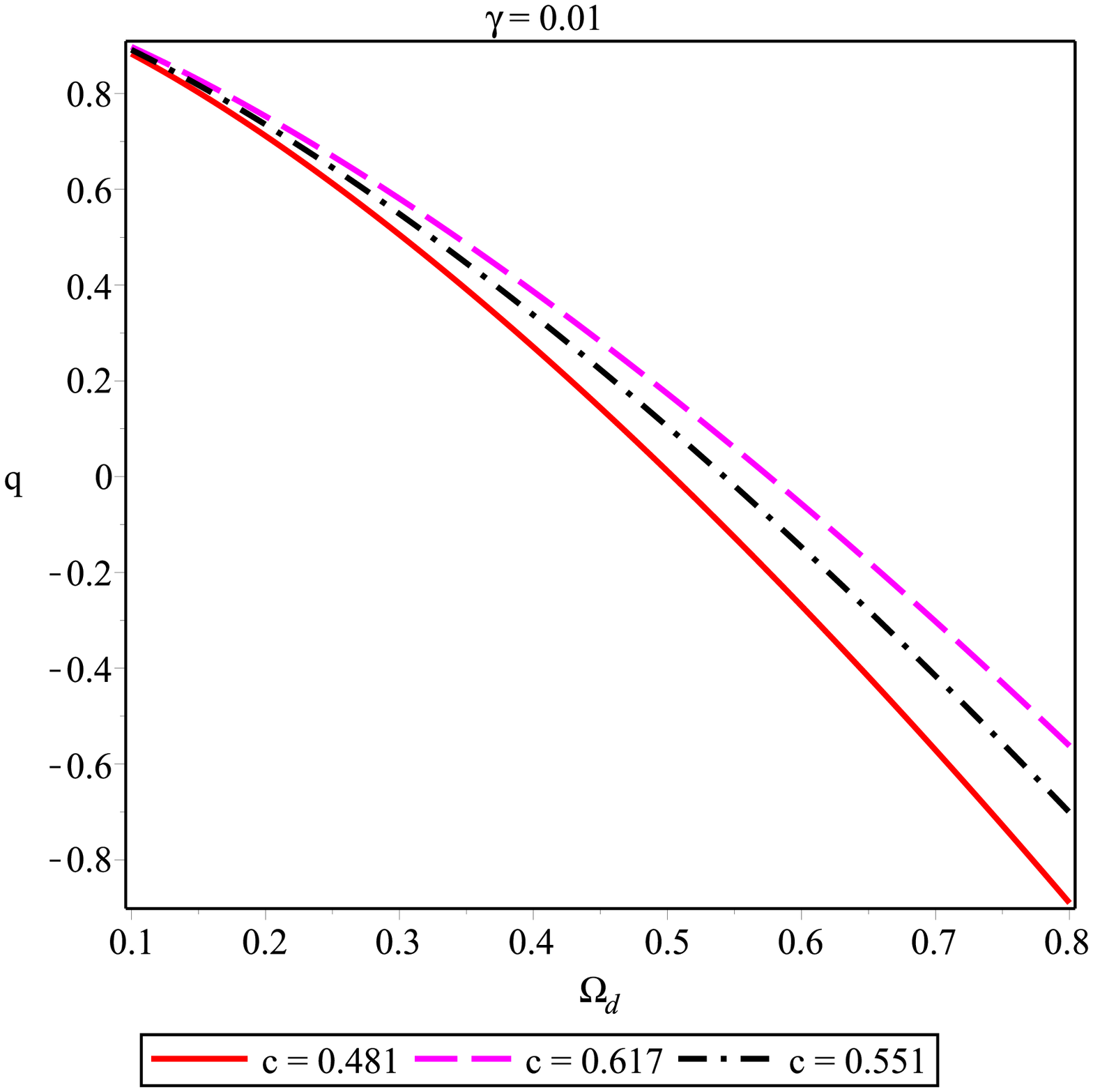}\\
Figure 5(b): This figure also represents the variation of the deceleration parameter with the dark energy density parameter for HDE at future event horizon for three different Planck data set.
\end{minipage}
\begin{minipage}{0.4\textwidth}
\includegraphics[width= 1.0\linewidth]{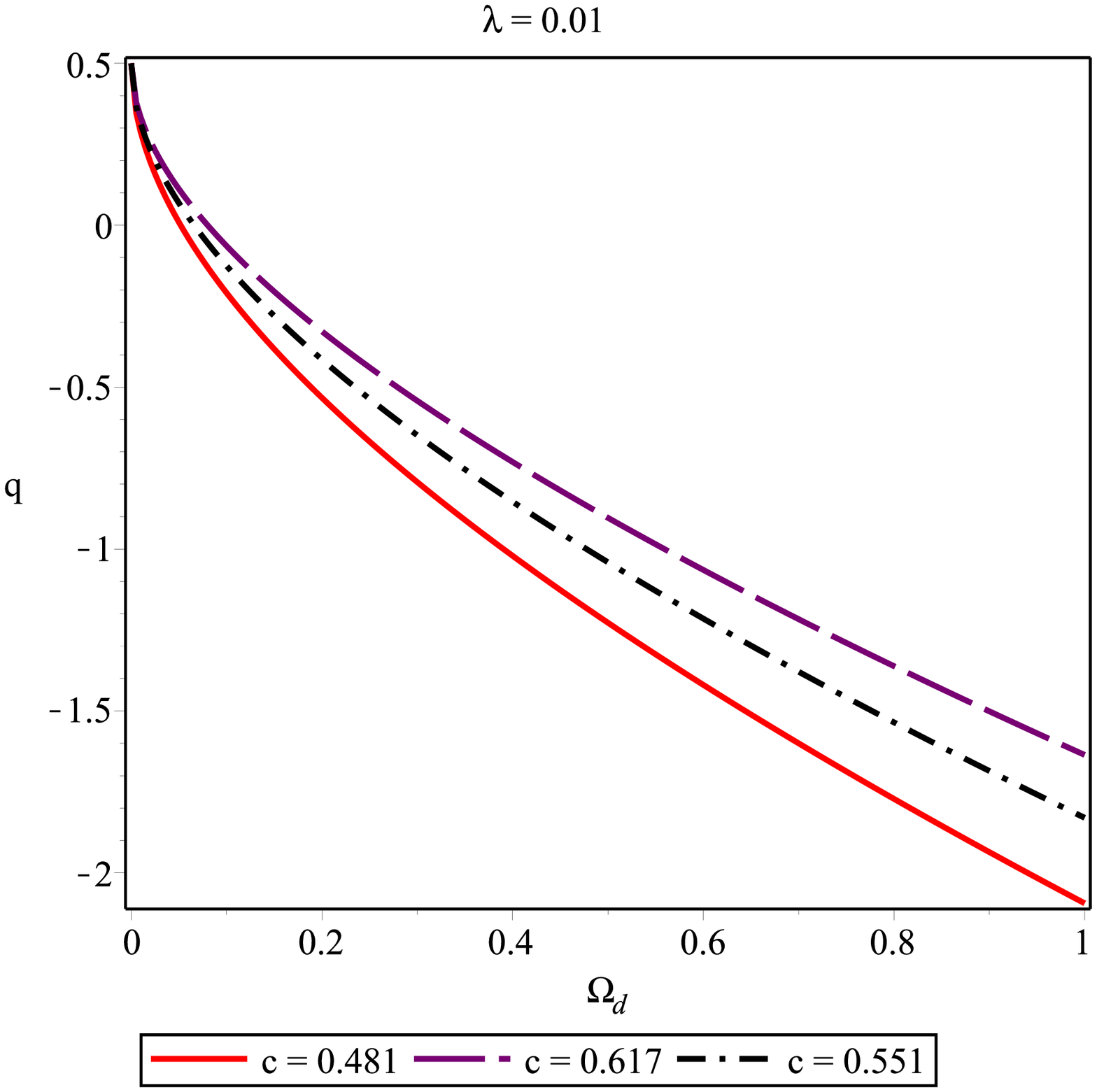}\\
Figure 5(c): This figure also represents the variation of the deceleration parameter with the dark energy density parameter for HDE at future event horizon for three different Planck data set.
\end{minipage}
\end{figure}

\begin{figure}
\begin{minipage}{0.4\textwidth}
\includegraphics[width= 1.0\linewidth]{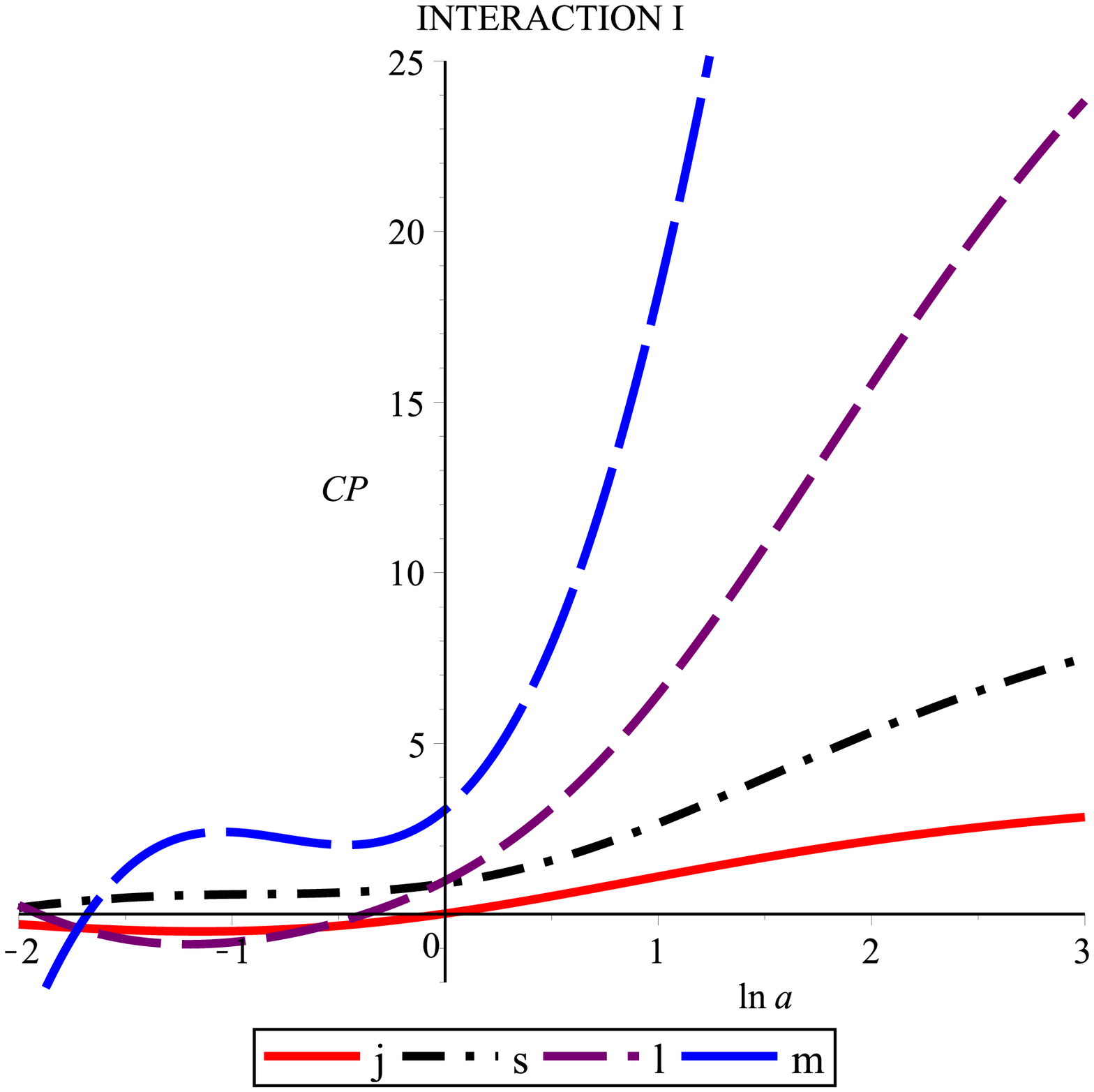}
Figure 6(a): The cosmographic parameters $j$, $s$, $l$ and $m$ are plotted for the holographic dark energy at Ricci scale for the first interaction term. We have taken $b^2= 0.01$.
\end{minipage}
\begin{minipage}{0.4\textwidth}
\includegraphics[width= 1.0\linewidth]{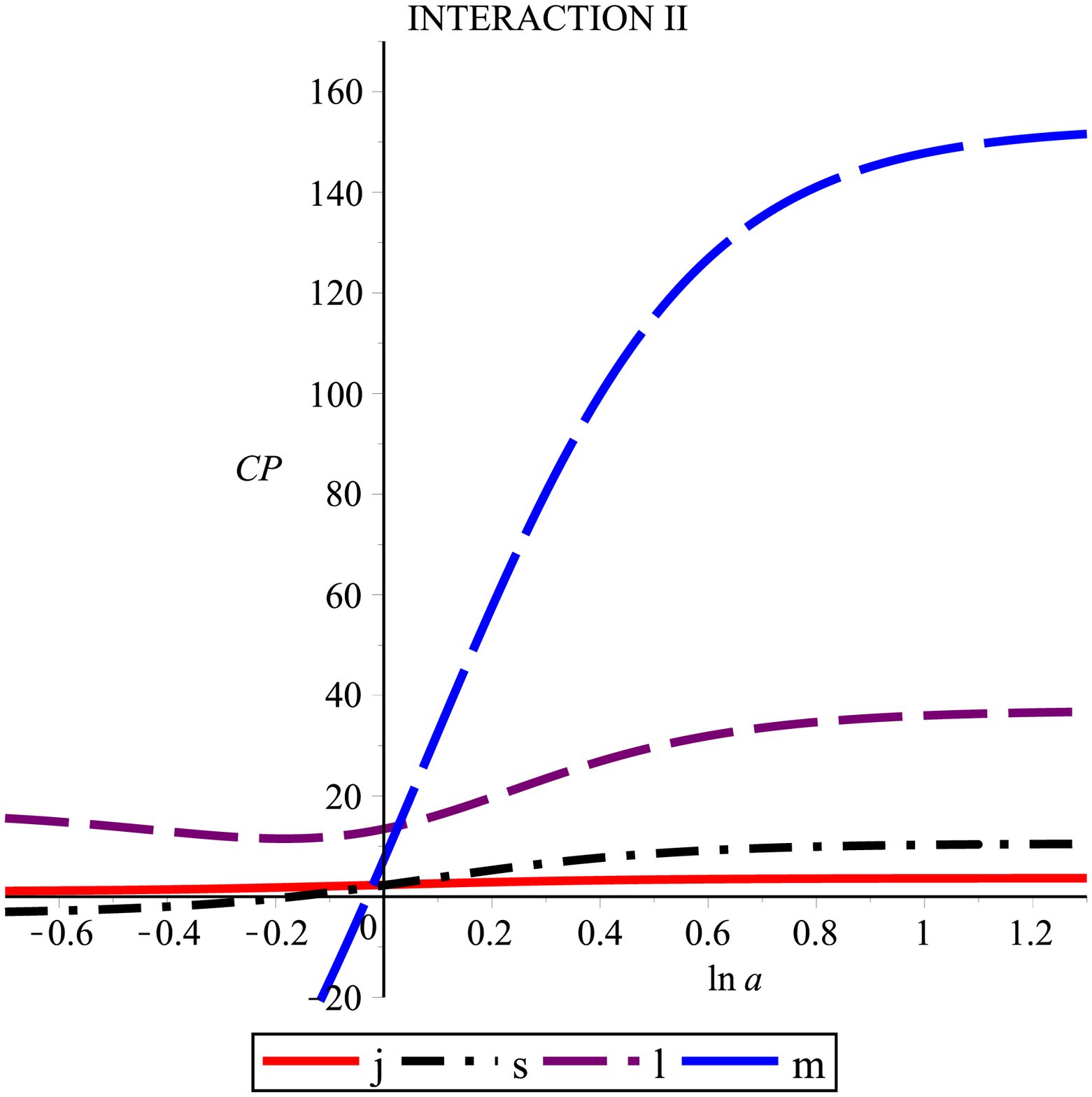}\\
Figure 6(b):  For the second interaction term, the cosmographic parameters $j$, $s$, $l$ and $m$ are plotted for the holographic dark energy at Ricci scale with $\gamma= 0.01$.
\end{minipage}
\begin{minipage}{0.4\textwidth}
\includegraphics[width= 1.0\linewidth]{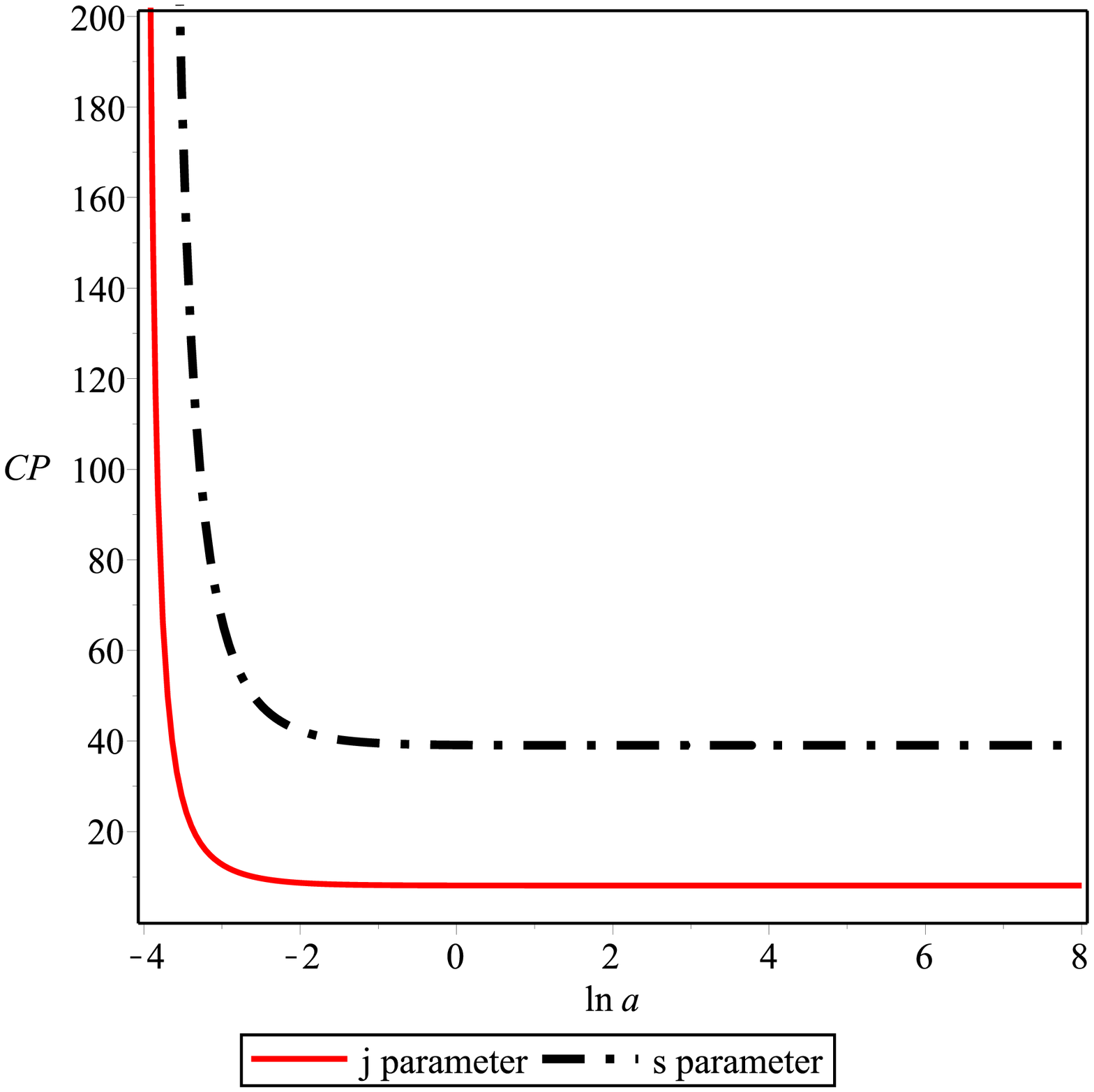}\\
Figure 6(c):  For the third interaction term, the above two graphs represent the $j$ and $s$ parameters for $\lambda= 0.01$
\end{minipage}
\begin{minipage}{0.4\textwidth}
\includegraphics[width= 1.0\linewidth]{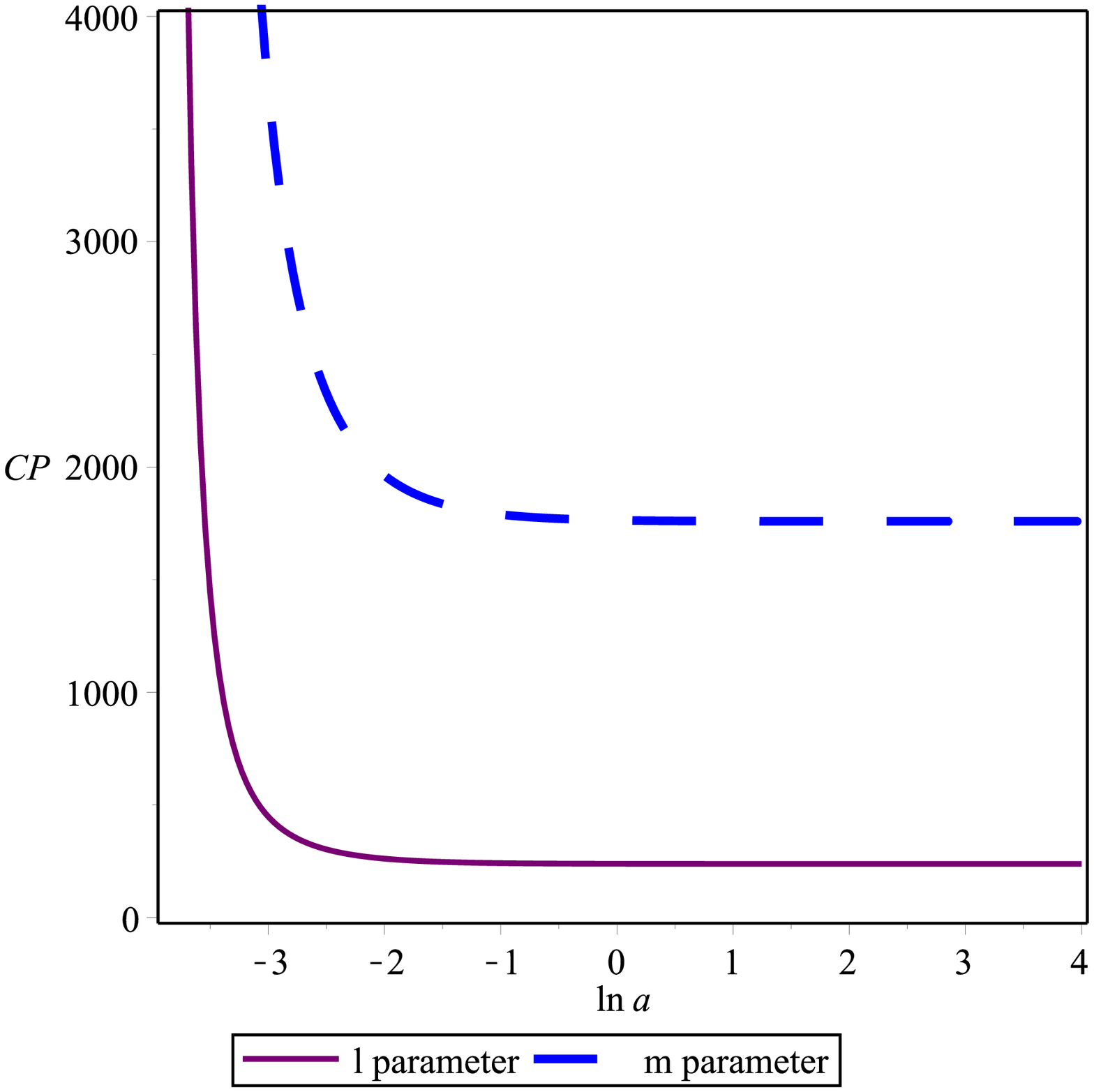}
Figure 6(d): These two graphs represent $l$ and $m$ parameters for the third interaction term with small coupling parameter $\lambda= 0.01$.
\end{minipage}
\end{figure}

\section{COSMOGRAPHY of Holographic dark energy}

To have a geometric view of the dark energy models, Sahni et al. \cite{Sahni1} first proposed two geometric parameters $r$ and $s$. The interesting fact regarding these geometric parameters are that, they filters the well supported dark energy models from infinitely many DE models and they are model independent. This geometric investigations on the DE models have been extended considering the Taylor series expansion of the scale factor about the present time. As a result we have model independent parameters $j$, $s$, $l$ and $m$ known as CP \cite{Visser1, Chakraborty1} and are defined as $$j= \frac{1}{aH^3} \frac{d^3 a}{dt^3},~~s= \frac{1}{aH^4} \frac{d^4 a}{dt^4},~~l= \frac{1}{aH^5}\frac{d^5 a}{dt^5},~~and~~m= \frac{1}{aH^6}\frac{d^6 a}{dt^6}.$$

Or, equivalently, in terms of the deceleration parameter, the CP are expressed as (remembering $x$= $ln a$)

$$q= 1+ \frac{3}{2}\Omega_m \omega_m -\frac{\Omega_d}{c^2},$$
$$j= -\frac{dq}{dx}+2(1+q)^2-3q-2,$$
$$s= \frac{dj}{dx}-(2+3q)j,$$
$$l= \frac{ds}{dx}-(3+4q)s,$$
\begin{equation}
m= \frac{dl}{dx}-(4+5q)l.
\end{equation}

Using the data set ``Planck+ WP+ Union 2.1+ BAO+ HST+ lensing" \cite{Li2}, we have graphically shown the cosmographic parameters $j$, $s$, $l$ and $m$ in Figures 6(a), 6(b) and 6(c)--6(d) for three different interactions, respectively, with very small coupling parameter only for HDE at Ricci length scale.\\\\

$~~~~${\bf Table I:} Planck data set (Ref. \cite{Li2}).\\\\
\begin{tabular}{|c|c|c|c|c|c|}
\hline Data set  & c  & $\Omega_ d$  & Theoretical $\Omega_d$  &~~~ $a_0$~~  & $b_0$ \\
\hline Planck+WP+BAO+lensing      & 0.498 & 0.738 & $\approx$ 0.73 & $\approx$ -0.4  & -0.5 --- -0.9\\
\hline Planck+WP+BAO+HST+lensing     & 0.481 & 0.755 & $\approx$ 0.75 & $\approx$ -0.4 & -0.5 --- -0.9\\
\hline Planck+WP+Union 2.1+lensing    & 0.617 & 0.679 &  $\approx$ 0.68  & $\approx$ -0.4 & -0.5 --- -0.9\\
\hline Planck+WP+Union 2.1+BAO+HST+lensing     & 0.551 & 0.724 & $\approx$ 0.72 & $\approx$ -0.4 & -0.5 --- -0.9 \\
\hline
\end{tabular}\\

\section{Summary and Conclusions}

In the present work, an explicit study of holographic dark energy (HDE) model has been done considering IR cut off as: (i) Ricci length scale, and (ii) future event horizon. Depending on different choices, the cosmological dynamics perform different behaviors. Exact analytic solution is possible for HDE with Ricci length scale while due to complicated form of the evolution equation, exact analytic solutions can not be obtained for HDE at future event horizon and cosmological parameters are determined as functions of the HDE density parameter only. Among three possible choices of the interactions, the cosmological parameters show a distinct features for the third type of interaction ({\it i.e.,} $Q= 3 \lambda H \rho_d$). The coupling parameter `$c$' in the HDE model is estimated from four Planck data set. We then discussed a cosmographic analysis only for HDE with Ricci length scale (due to its simple form). The graphs of different geometric CP show that they are negative in early phases of the evolution and then they gradually increase and become $+$ve for the first two interactions while for the third interaction, the CP are positive throughout and they decrease sharply from high value at the early epoch and then they become constant.\\

Also, from the figures, it is evident that the cosmological parameters for the first two type of interactions is well within the observed range of the parameters but for the third choice of interaction, the said parameters do not match with the observations. So, based on the cosmographic analysis, we may conclude that for HDE the first two type of interactions are reasonable and the third one should be rejected.\\

Finally, the present interacting HDE model shows a counter example for the `no-go theorem' \cite{yfcai2, Xia1} (particularly from Figures 2, 4 (a) and 4 (c)) and it is interesting to mention that the numerical plots in Figure 2, where the equation of state parameter crosses the cosmological constant boundary $\omega= -1$, have similarities to a dark energy model of a spinor field discussed by the authors in Ref. \cite{yfcai3}.\\

$~~~~~~~~~~~~~~~~~~~~~~~~~~~~~~~~~~~~~~~~~~~~~~~${\bf Acknowledgments}\\

SP thanks CSIR, Govt. of India for financial support through SRF scheme (File No: 09/096 (0749)/2012-EMR-I). Author SC thanks UGC-DSR programme at the Department of Mathematics, Jadavpur University. Finally, the authors are grateful to the anonymous referee whose valuable and encouraging suggestions helped us to improve the manuscript considerably.

\end{document}